\documentclass[prapplied,twocolumn,superscriptaddress,notitlepage]{revtex4-2}

\usepackage{soul}
\usepackage{amsthm}
\usepackage{amsmath}
\usepackage{amssymb}
\usepackage{graphicx}
\usepackage{epstopdf}
\usepackage{subfig}
\usepackage{multirow}
\usepackage{url}
\usepackage{float}
\usepackage{bm}
\usepackage{ifthen}
\usepackage[usenames,dvipsnames]{color}
\usepackage{mathrsfs}
\usepackage[colorlinks=false,citecolor=blue,urlcolor=black]{hyperref}
\usepackage[absolute]{textpos}
\usepackage{xcolor}

\newcommand{\sket}[1]{{\ensuremath{\lvert#1\rangle}}}
\newcommand{\lket}[1]{{\ensuremath{\left\lvert#1\right\rangle}}}
\newcommand{\ket}[1]{\if@display\lket{#1}\else\sket{#1}\fi}

\newcommand{\sbra}[1]{{\ensuremath{\langle#1\rvert}}}
\newcommand{\lbra}[1]{{\ensuremath{\left\langle#1\right\rvert}}}
\newcommand{\bra}[1]{\if@display\lbra{#1}\else\sbra{#1}\fi}

\newcommand{\sbraket}[2]{{\ensuremath{\langle#1\rvert#2\rangle}}}
\newcommand{\lbraket}[2]{{\ensuremath{\left\langle#1\!\left\rvert\vphantom{#1}#2\right.\!\right\rangle}}}
\newcommand{\braket}[2]{\if@display\lbraket{#1}{#2}\else\sbraket{#1}{#2}\fi}

\newcommand{\sketbra}[2]{{\ensuremath{\lvert #1\rangle\!\langle #2\rvert}}}
\newcommand{\lketbra}[2]{{\ensuremath{\left\lvert #1\right\rangle\!\!\left\langle #2\right\rvert}}}
\newcommand{\ketbra}[2]{\if@display\lketbra{#1}{#2}\else\sketbra{#1}{#2}\fi}

\usepackage{color}
\newcommand{\mbb}{\mathbb}
\newcommand{\mc}{\mathcal}
\newcommand{\tr}{\text{Tr}}
\newcommand{\op}[2]{|#1\rangle\langle #2|}
\newcommand{\steer}{\text{Steer}}

\definecolor{cool_green}{rgb}{0.0, 0.5, 0.0}

\graphicspath{{./images/},{./}}

\begin{document}
\title{Entanglement Verification of Hyperentangled Photon Pairs}
\author{Christopher K. Zeitler}
\affiliation{Illinois Quantum Information Science and Technology Center, University of Illinois Urbana-Champaign, Urbana, IL 61801}
\affiliation{Department of Physics, University of Illinois Urbana-Champaign, Urbana, IL 61801}
\author{Joseph C. Chapman}
\affiliation{Illinois Quantum Information Science and Technology Center, University of Illinois Urbana-Champaign, Urbana, IL 61801}
\affiliation{Department of Physics, University of Illinois Urbana-Champaign, Urbana, IL 61801}
\affiliation{Oak Ridge National Laboratory, Oak Ridge, TN 37831}
\author{Eric Chitambar}
\affiliation{Illinois Quantum Information Science and Technology Center, University of Illinois Urbana-Champaign, Urbana, IL 61801}
\affiliation{Department of Electrical \& Computer Engineering, University of Illinois Urbana-Champaign, Urbana, IL 61801}
\author{Paul G. Kwiat}
\email{kwiat@illinois.edu}
\affiliation{Illinois Quantum Information Science and Technology Center, University of Illinois Urbana-Champaign, Urbana, IL 61801}
\affiliation{Department of Physics, University of Illinois Urbana-Champaign, Urbana, IL 61801}

\begin{abstract}
We experimentally investigate the properties of hyperentangled states displaying simultaneous entanglement in multiple degrees of freedom, and find that Bell tests beyond the standard Clauser, Horne, Shimony, Holt inequality can reveal a higher-dimensional nature in a device-independent way. Specifically, we show that hyperentangled states possess more than just simultaneous entanglement in separate degrees of freedom but also entanglement in a higher dimensional Hilbert space. We also verify the steerability of hyperentangled quantum states by steering different photonic degrees of freedom.
\end{abstract}
\maketitle

\begin{textblock}{13.3}(1.4,15)\noindent\fontsize{7}{7}\selectfont\textcolor{black!30}{This manuscript has been co-authored by UT-Battelle, LLC, under contract DE-AC05-00OR22725 with the US Department of Energy (DOE). The US government retains and the publisher, by accepting the article for publication, acknowledges that the US government retains a nonexclusive, paid-up, irrevocable, worldwide license to publish or reproduce the published form of this manuscript, or allow others to do so, for US government purposes. DOE will provide public access to these results of federally sponsored research in accordance with the DOE Public Access Plan (http://energy.gov/downloads/doe-public-access-plan).}\end{textblock}
\section{Introduction}
Entanglement---evincing nonlocal correlations that exceed what is possible according to any local realistic model, i.e., local hidden variables---is at the very foundation of quantum mechanics and underlies much of the new quantum information revolution. 
In the 1960s, John Bell developed a test to distinguish such hidden-variable theories from quantum mechanical ones~\cite{PhysicsPhysiqueFizika.1.195} by specifying a quantity that had different maximal bounds in the two models.   Since their advent, Bell tests have been a focus of fundamental research in physics, providing a means to demonstrate the nonlocal effects present in quantum mechanics~\cite{PhysRevLett.23.880}, verify the presence of entanglement~\cite{RevModPhys.86.419}, and even explore the limits of ultra-nonlocal theories, which can predict correlations stronger than those allowed by standard quantum mechanics~\cite{PhysRevX.5.041052}. Other techniques, such as quantum steering~\cite{schrodinger_1935,
schrodinger_1936,
PhysRevLett.98.140402,RevModPhys.92.015001}, expand the applicability of entanglement verification to a wider set of scenarios with differing assumptions.  Initially, these tests of nonlocality were conceived of as ``thought experiments'' that revealed unexpected (or to some, illogical) features of quantum mechanics; however, repeated experimental verification of the correlations that are the hallmark of entangled states has left little doubt that ``spooky action at a distance'' is a part of reality.  The refinement of these measurement techniques culminated in a trio of ``loophole-free'' tests of nonlocality using Bell inequalities, providing compelling evidence that Nature is truly nonlocal~\cite{PhysRevLett.115.250402,PhysRevLett.115.250401,Hensen2015}. Meanwhile, loophole-free versions of quantum steering have also been reported~\cite{Wittmann_2012}. The fundamental importance of such experiments testing the features of nonlocality was wonderfully highlighted by the recent Nobel Prize in Physics awarded to pioneers in the field John F. Clauser, Alain Aspect, and Anton Zeilinger~\cite{nobelprize.org_2022}.

Now that the original purpose of Bell tests, providing a measurable criteria for separating local and nonlocal theories, has been largely fulfilled, a new era for Bell tests is unfolding, in which they are used as tools for probing and verifying the properties of quantum states. Most recently, Bell tests have emerged as a resource for generating provably random number strings~\cite{PhysRevLett.94.050503,PhysRevLett.111.130406}, and as a means to ensure cryptographic security in ``device-independent'' quantum-key-distribution protocols without needing to trust all the devices \cite{PhysRevLett.108.130503, PhysRevLett.98.230501}.
The ability to draw conclusions about a measured quantum state without needing to trust the devices used to make the measurement or prepare the state is a defining feature of Bell tests and can be used to distinguish them from other methods of verifying entanglement.  Quantum steering represents an intermediate case~\cite{RevModPhys.92.015001}, in which one of the measurement devices must be trusted, while no assumptions are made about the second measurement device.  

Most Bell tests to date have used some version of the Clauser, Horne, Shimony, Holt (CHSH) inequality~\cite{PhysRevLett.23.880} applied to qubits, but the space of possible tests and states to test is much larger~\cite{PhysRevX.5.041052}, e.g., including bipartite systems of higher dimensionality, which has been partially explored~\cite{Dada2011,dada2011bell,Lo2016,Kues2017}.  In photonic systems, there have been experimental demonstrations of quantum steering for both qubit-entangled states~\cite{Saunders-2010a, Bennet-2012a, Yang:21} and higher-dimensional ones~\cite{PhysRevLett.120.030401,Lee:18,PhysRevLett.126.200404,PhysRevLett.128.240402,srivastav2022noise}.  However, in all such previous steering demonstrations, the entanglement has been shared on a single degree of freedom (DOF) of the photon (e.g., polarization, frequency, position-momentum, etc.). Using multiple DOFs, a single photon is able to carry more than just a qubit of quantum information, and when two photons are entangled in more than one DOF, higher-dimensional entanglement can be realized, a phenomenon known as hyperentanglement~\cite{Kwiat-1997a,PhysRevLett.95.260501}.  The use of multiple photonic DOFs and hyperentangled states has already shown advantages in tasks like state discrimination~\cite{PhysRevA.84.022340}, entanglement distribution~\cite{Piparo-2019a} and distillation~\cite{PhysRevLett.127.040506}, teleportation~\cite{Graham2015}, and quantum error correction~\cite{Piparo-2020a}. Thus, hyperentanglement is a promising candidate for achieving more efficient and noise-robust quantum communication.

In this paper, we investigate nonlocality tests on hyperentangled quantum states.  By coherently controlling two DOFs on each photon, we can certify genuine higher-dimensional entanglement, a task that can otherwise be quite demanding to achieve experimentally.  After describing our experimental setup, we discuss results of a higher dimensional Bell inequality, where we certify entanglement with dimensionality larger than just two qubits. Polarization and path hyperentanglement has been used previously in several tests of non-locality~\cite{PhysRevLett.97.140407,PhysRevA.85.032107}. Our source uses polarization and time-bin hyperentanglement for optimal applicability to the space-to-earth channel~\cite{chapman2020time,chapman2019hyperentangled}. We also optimize the choice of higher-dimensional inequality and characterize the violation as a function of added decoherence. Finally, we display quantum steering of our hyperentangled two-particle state, also as a function of added decoherence.  To our knowledge, this is the first steering demonstration across multiple DOFs of an entangled single photon.
 
\section{Experimental Methods}
\begin{figure}
\centerline{\includegraphics[width=1\columnwidth]{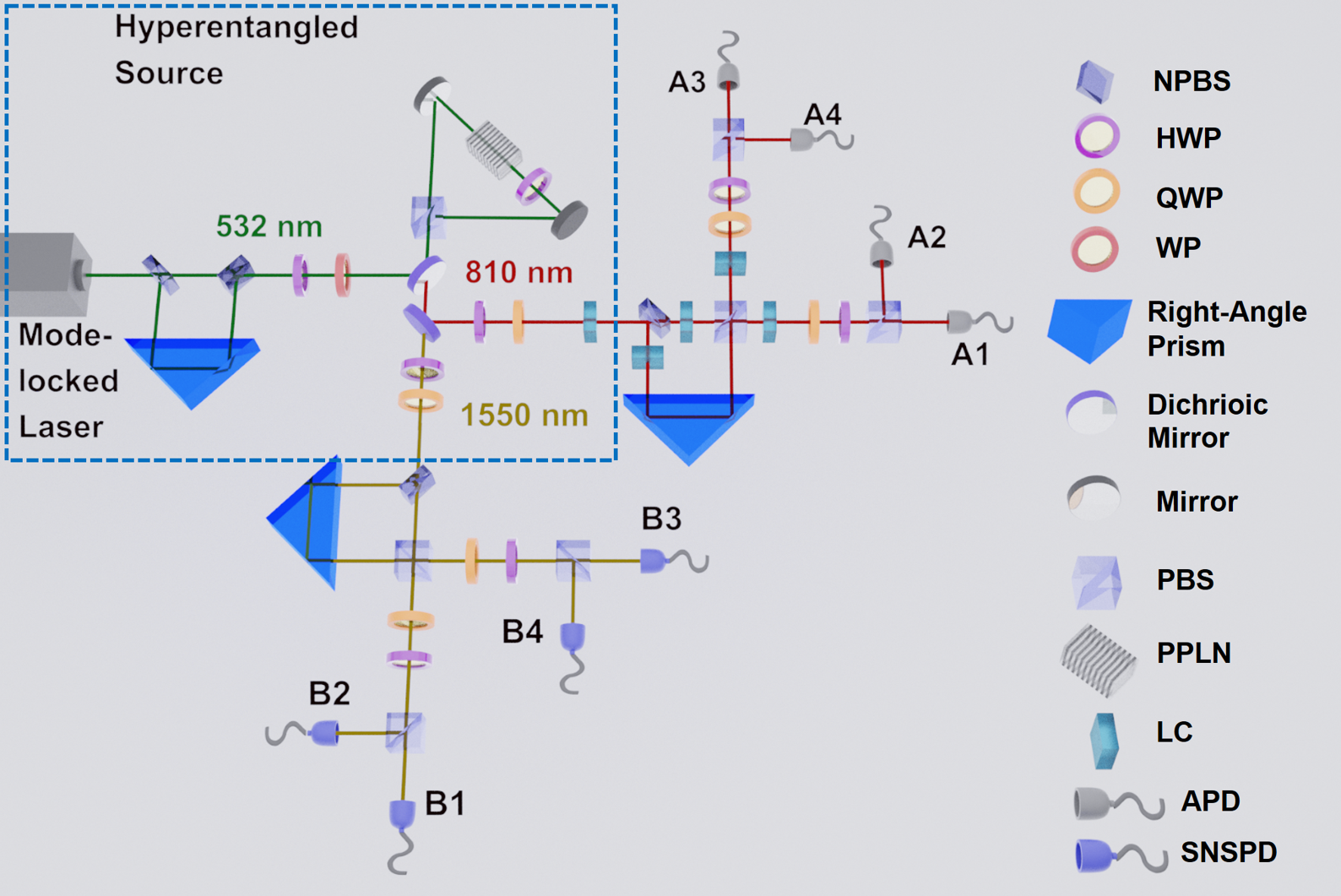}}
\caption{Schematic of the hyperentangled photon system. The pump is prepared in a superposition of timing modes so that, when it passes through the Sagnac polarization entanglement source, the output is entangled in both polarization and time-bin.  The photons are then separated by a dichroic mirror for further analysis. The measurement system combines two standard polarization analysis systems with a polarization-dependent unbalanced interferometer. This leads to coupling between the polarization and timing modes so that timing measurements can be controlled with polarization optics.  This measurement system is duplicated for both photons, with the 810-nm photons being detected by APDs and the 1550-nm photons being detected by SNSPDs.  The liquid crystals in the 810-nm measurement system are used to tune the phases in the generated state and of the measurement system. HWP $\equiv$ half-wave plate. QWP $\equiv$ quarter-wave plate. WP $\equiv$ wave plate. PBS $\equiv$ polarizing beam splitter. PPLN $\equiv$ periodically poled lithium niobate. LC $\equiv$ liquid crystal. APD $\equiv$ avalanche photodiode. SNSPD $\equiv$ superconducting-nanowire single-photon detector.}
\label{setupfig}
\end{figure}

The time-bin and polarization hyperentangled photon pair source, shown in Fig.~\ref{setupfig}, is driven by a 532-nm pulsed laser  (Spectra Physics Vanguard 2.5W 355 laser, frequency doubled from 1064 nm) with an 80-MHz repetition rate.  The pump laser is sent through an unbalanced Mach-Zehnder interferometer to put each pulse into a superposition of an early and a late time bin, separated by 2.4 ns, large compared to the 7-ps pump-pulse duration.  After the interferometer, the polarization of the pump beam is prepared using wave plates, after which the pump enters a polarizing Sagnac interferometer~\cite{Sagnac1,Sagnac2}. The Sagnac interferometer contains a periodically poled lithium niobate (PPLN) crystal (poling period 7.5 $\mu$m), inside which the pump undergoes type-0 phase-matched spontaneous parametric downconversion. The horizontal (vertical) component of the pump traverses the Sagnac loop clockwise (counterclockwise) to produce a pair of horizontally polarized photons with wavelengths 810 nm and 1550 nm. The interferometer also contains a Fresnel rhomb, acting as a broadband half-wave plate, causing the output of the clockwise path to be vertically polarized, while the counter-clockwise path's output is horizontally polarized (because the counter-clockwise-propagating vertical-polarization pump was converted to horizontal polarization before the PPLN crystal), leading to polarization entanglement. A 4.1-cm piece of calcite (Thorlabs BD40) in the Sagnac (after the transmitted port of the PBS) acts as temporal compensation, to counteract the wavelength-dependent delay of the Fresnel rhomb. After the Sagnac, in the 1550-nm path, there is another small piece of calcite (0.5-mm long) to fix the delay mismatch more precisely between the horizontal and vertical polarizations exiting the Sagnac. Due to dispersion, only the 1550-nm side needed extra temporal compensation outside of the Sagnac. For more information on this source, see Ref.~\cite{chapman2020time}. 

In general, there will be relative phases between the two polarization modes, the two-timing modes, and the polarization and timing modes
\begin{align}
 	\ket{\psi}&=\frac{1}{2}(\ket{HH}+e^{i\phi_p}\ket{VV})\otimes(\ket{t_1t_1}+e^{i\phi_t}\ket{t_2t_2})\\
 	&=\frac{1}{2}(\ket{00}+e^{i\phi_t}\ket{11}+e^{i\phi_p}\ket{22}+e^{i(\phi_t+\phi_p)}\ket{33}),
\end{align}
where in our system we assign $\ket{0}\equiv \ket{Ht_1}, \ket{1}\equiv \ket{Ht_2}, \ket{2}\equiv \ket{Vt_1}$, and $\ket{3}\equiv \ket{Vt_2}$.
 In our experiments, these phases are set using liquid crystal elements acting on the 810-nm photons. 

\begin{figure}
    \centering
    \includegraphics[width=1\columnwidth]{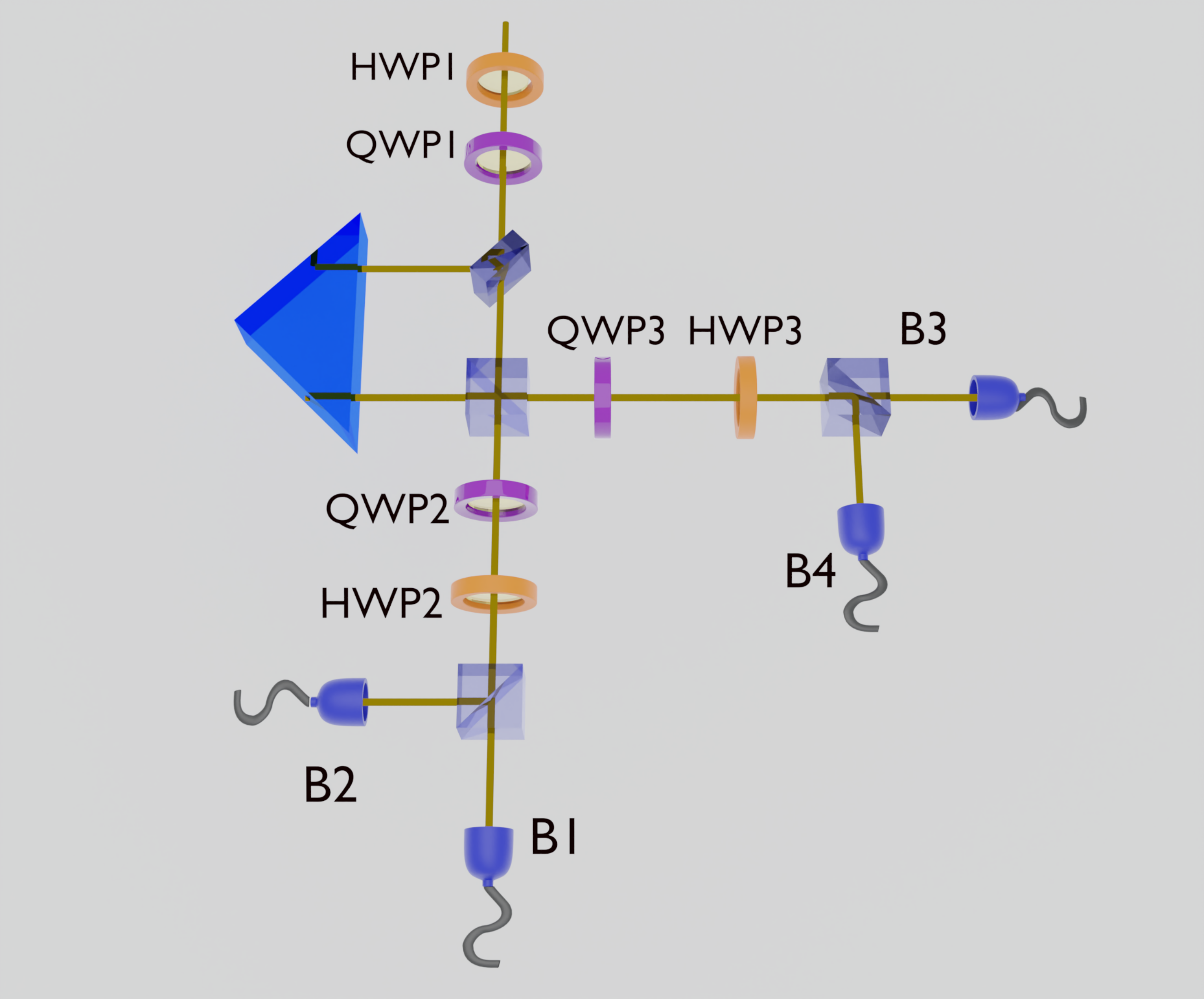}
    \caption{Detail of analyzer interferometer, with photons entrance from the top. Here we show the naming convention of the wave plates used as part of the measurement system.}
    \label{fig:wpcloseup}
\end{figure}

After exiting the Sagnac interferometer, the downconversion photons are separated from the pump and each other using dichroic mirrors, before being routed to the measurement system.  The measurement system is designed so that both the time and polarization measurements can be carried out using polarization optics.  This requires a coupling between the polarization and time modes, achieved using another unbalanced Mach-Zehnder interferometer that contains a final PBS (see Fig.~\ref{fig:wpcloseup}): the PBS couples the timing and polarization modes, allowing the analysis timing mode to be effectively controlled using waveplates.

In Fig.~\ref{fig:wpcloseup}, we denote HWP1 and QWP1 as the pair of waveplates before the analyzer interferometer; HWP2 and QWP2 as the pair of waveplates after the interferometer, but before the PBS in front of Detectors 1 and 2; and HWP3 and QWP3 as the pair of waveplates after the interferometer, but before the PBS in front of Detectors 3 and 4. For example, with QWP2 and HWP2 at $0^{\circ}$ (with respect to the horizontal), Detector 1 will project onto the first timing mode, with the polarization mode set by QWP1 and HWP1; with HWP1 and QWP1 at $0^\circ$, for instance, the measurement state for Detector 1 is $\bra{Ht_1}$ or, with HWP1 at $22.5^\circ$, $\bra{Dt_1}$. With HWP2 at $22.5^\circ$, Detector 1 will project onto an equal superposition of both timing modes with orthogonal polarizations; for example, with QWP1 and HWP1 at $0^\circ$, the measurement state will be $(\bra{Ht_1}+\bra{ Vt_2})/\sqrt{2}$). In this way, the relevant amplitude and phase between the measured timing modes can be controlled with HWP2 and QWP2, while the measured polarization state is controlled by HWP1 and QWP1.  

Due to the probabilistic nature of the first non-polarizing beamsplitter in this interferometer, three possible measurement time bins are generated.  The early (late) time-bin arises from photon pairs created by the pump pulse that went through the short (long) arm of the pump interferometer and were subsequently sorted into the short (long) arm of the analysis interferometer; the middle time-bin corresponds to photons that took the short path in one interferometer and the long path in the other.  In this work, we post-select on this middle time-bin~\footnote{A polarization-independent optical switch~\cite{Trentthesis,10.1117/12.2537081} could be used to always route the early (late) photons to the long (short) paths, leading to a 4 times enhancement in the interference coincidence terms.}. Because we have matched the path length differences (long path minus short) in the two interferometers, these ``short-long'' and ``long-short'' processes are indistinguishable and, therefore, can interfere. To preserve the relative phase of the time modes, it is necessary to actively stabilize these measurement interferometers to match the pump interferometer. This active stabilization is carried out by sending a portions of the pump laser backwards through each measurement interferometers and measuring its intensity on photodiodes (Thorlabs PDA36A)~\cite{chapman2020time}. This feedback signal is then used to adjust the length of the long arm of the measurement interferometers using a piezoelectric element (Thorlabs AE0505D16F with Thorlabs TPZ001 driver) to vary the exact positions of the mirrors in the long arm; this discrete-time stabilization system has an output rate of about 100 Hz and yields an average phase stability of about $3^\circ$~\cite{chapman2020time}.  

The 810-nm photons are detected using four silicon avalanche photodiodes (Excelitas SPCM-AQ4C) with a measured detection efficiency of about $45\%$ at 810 nm and a timing jitter of about 600 ps. The 1550-nm photons are detected with four superconducting nanowire single-photon detectors (SNSPDs) with an efficiency of about $80\%$ and a timing jitter of less than 100 ps. The outputs of these detectors are then sent to a time-bin discrimination circuit~\cite{chapman2019hyperentangled} and then to fast time-tagging electronics (UQDevices UDQ-Logic-16), yielding a net coincidence timing resolution of about 700 ps. The resulting time tags are processed to determine coincidence events corresponding to detections in the middle time-bin, which is then easily distinguished from the early and late time bins at $\pm$2400 ps, using the sorting capability of the time-bin discrimination circuit. Unfortunately, for steering measurements in Fig.~\ref{steerplot} with visibility below 0.6, the time-bin discrimination circuit channel for Detector A4 was not functional, so each measurement was repeated two times---the second time a HWP was used to direct the A4 events to the functional Detector A3. This does not present an issue because our measurement system is not attempting to close any loopholes and our system was stable over longer than the time to take multiple measurements.

\section{Verifying Higher-Dimensional Hyperentanglement}
Bell inequalities can be characterized by the number of bases measured and the number of outcomes of those measurements;  in general, these values can be different for the two parties making measurements, so bipartite inequalities require four parameters to be classified~\cite{4Dineq}. The CHSH Bell inequality uses two measurement bases and two measurement outcomes on each side, and can be referred to as $I_{2222}$. For that particular set of parameters, there is only one possible inequality (ignoring trivial permutations), and this cannot distinguish between entangled qubits and higher dimensional entanglement. In contrast, for bases and outcomes above 2 there can be multiple inequalities for a given set of parameters.  Here, we verify the higher-dimensional entanglement of the hyperentangled time-bin and polarization state by making a Bell inequality measurement that can produce different violations depending on state dimensionality.  Specifically, we focus on the symmetric case of using four measurement bases and two measurement outcomes for both photons in the pair ($I_{4422}$), because some of these inequalities display different quantum bounds for qubits and qutrits~\cite{P_l_2008}. With these parameters, there are at least 27 known inequalities~\cite{4Dineq}. For completeness, we also perform standard CHSH inequality measurements on pairs of photons entangled in polarization and separate measurements on pairs of photons entangled in time-bin modes but these alone are insufficient to say the state possesses higher dimensional entanglement. See Appendix~\ref{sec:CHSH} for a description of CHSH measurements and results.

Both observed CHSH Bell parameters nevertheless indicate that the source was entangled in each degree of freedom separately. Notably, however, measuring a CHSH inequality of this type \textit{cannot} be used to infer that the state was entangled in both degrees of freedom simultaneously, because formally the CHSH inequality can always exhibit a violation even if the source is only entangled in one degree of freedom; this is true even if the measurement settings are chosen to depend on both polarization and time mode, instead of measuring each degree of freedom separately.  Unless an assumption is made that the states Alice and Bob projected onto were indeed the ones they intended (or announced they would measure), it is not possible to conclude that the state was hyperentangled from just a CHSH Bell measurement.  However, an important feature of Bell tests is that they need not rely on such an assumption about what measurements were actually carried out; because the Bell test can be interpreted solely as a mathematical game on boxes with local realistic constraints~\cite{RevModPhys.86.419}, it does not depend on what quantum states might violate it. Consider the scenario in which Alice and Bob both try to measure in a hybrid basis between polarization and time-bin that should only yield a violation for truly hyperentangled states, but problems in their measurement devices make their results insensitive to timing information, so the actual bases used depend only on polarization.  In this case, a state only entangled in polarization would lead Alice and Bob to \textit{incorrectly} conclude that they shared a hyperentangled state, because their assumption about what measurement basis they used was violated.  The synthesis of a Bell parameter and any conclusions drawn from it rely only on the correlations between Alice and Bob's results, and in this way, the CHSH inequality cannot provide information about a state's dimensionality. To provide dimensionality information via Bell tests, it is therefore critical that the conclusion not depend on trust of the measurement devices used.

In order to select an optimal higher-dimensional inequality to measure using our experimental system (which cannot measure all states), we carried out numerical simulations of the maximum Bell parameter attainable with our system for each of the known 27 inequalities (see Appendix~\ref{sec:BIAO} for short discussion of limitations on inequality choice), using a maximally entangled ququart state as an input:
\begin{equation}
\ket{\Psi_4}=\frac{1}{2}(\ket{00}+\ket{11}+\ket{22}+\ket{33}),
\label{psi4}
\end{equation}
where in our system we assign $\ket{0}\equiv \ket{Ht_1}, \ket{1}\equiv \ket{Ht_2}, \ket{2}\equiv \ket{Vt_1}$, and $\ket{3}\equiv \ket{Vt_2}$. Based on this analysis, we chose to measure $I^{18}_{4422}$ which is the inequality with the largest difference between our system's violation (constrained by our available measurements) using entangled ququarts and the theoretic bound using entangled qubits. With optimized wave-plate settings (see Appendix~\ref{sec:BIAO} for optimal settings), our system can theoretically achieve a Bell parameter of 0.46, while a local model is limited to a value of 0, and a qubit-entangled state is limited to 0.18~\cite{4Dineq,P_l_2008}. Note that $I^{18}_{4422}$ is also the $I_{4422}$ inequality that has the largest separation between qubit and qutrit performance for any allowed measurement basis~\cite{PhysRevLett.110.150501}; however, our system cannot create arbitrary measurements, so it is not able to reach the maximum quantum bound of 0.64~\cite{P_l_2008}.
 
We measured this $I^{18}_{4422}$ inequality using our best approximation to the maximally entangled ququart state given in Eq.~\eqref{psi4}. With optimal source tuning, we observed a parameter of $I^{18}_{4422}=0.45\pm0.03$ after measuring 85,000 coincidence events (summed over all measurement basis combinations), very close to the expected maximum value given our measurement limitations, proving that the dimensionality of the state must be larger than that of a qubit-entangled state. Since the state in Eq.~\ref{psi4} is actually composed of four-dimensional subsystems, one may wonder whether our data can verify this; unfortunately, as shown in Ref.~\cite{P_l_2008}, the $I^{18}_{4422}$ inequality is already saturated by entangled qutrits, i.e., it cannot distinguish between higher dimensional systems.

\begin{figure}
	\centerline{\includegraphics[width=1\columnwidth]{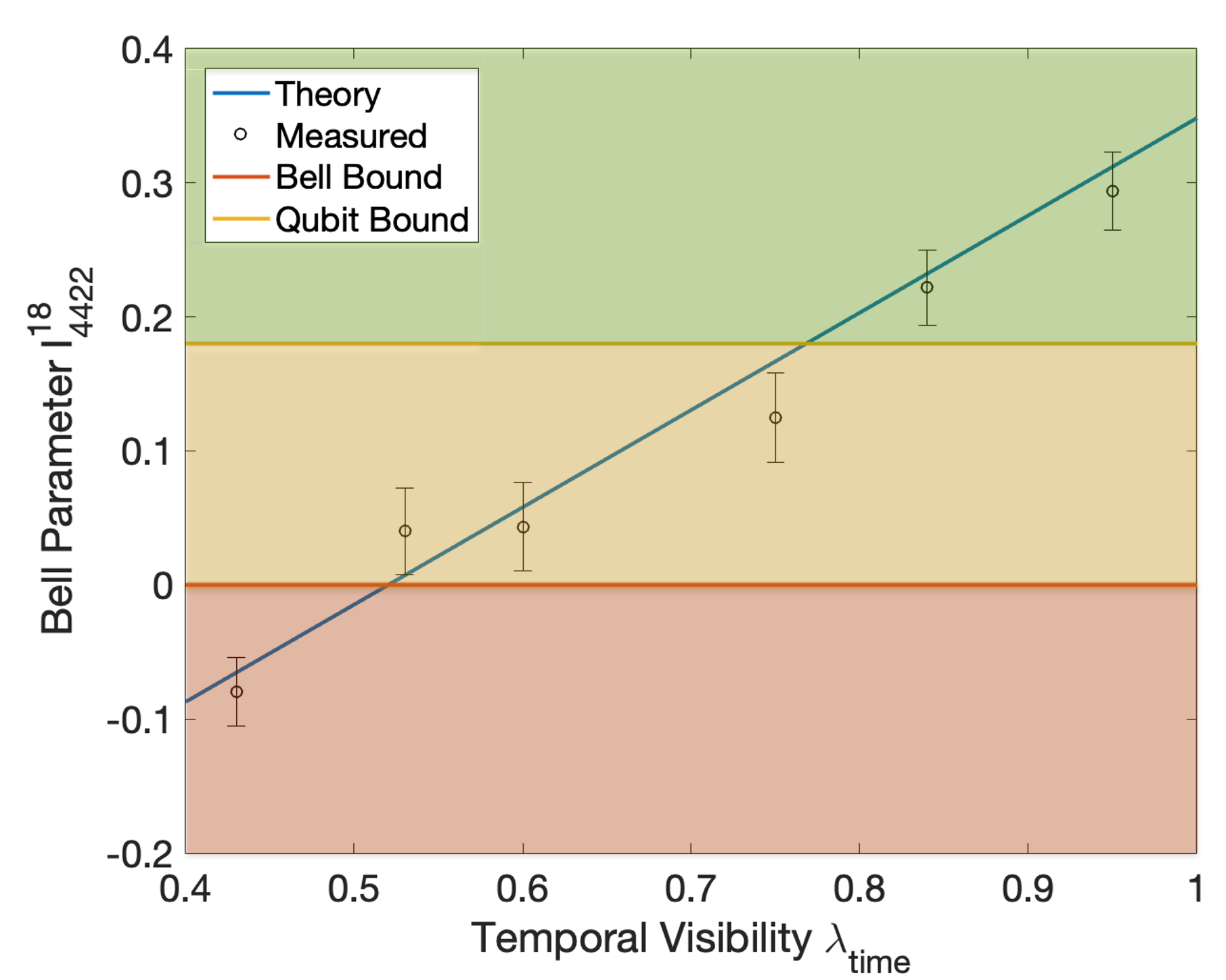}}
	\caption{Measured $I^{18}_{4422}$ values showing the Bell parameter visibility dependence. The visibility of the temporal qubit was changed by unbalancing the measurement interferometer's path length difference relative to the pump.  In the bottom red region, there is no evidence of nonlocality. In the yellow region, the state is nonlocal but could be a pair of entangled qubits. In the top green region, the state is nonlocal and must have a dimensionality higher than that of a pair of qubits, i.e., at least a pair of qutrits.}
	\label{visdep}
\end{figure}

We next investigated the robustness of this Bell parameter to state imperfections by reducing the amount of entanglement in the time-bin degree of freedom. This leads to a state of the form 
\begin{widetext}
\begin{equation}
\rho(\lambda_{pol},\lambda_{time})=(\tfrac{\lambda_{pol}}{2}\ket{\phi^+_{p}}\bra{\phi^+_{p}}+\tfrac{1-\lambda_{pol}}{2}(\ket{H}\bra{H}^{\otimes 2}+\ket{V}\bra{V}^{\otimes 2}))\otimes (\tfrac{\lambda_{time}}{2}\ket{\phi^+_{t}}\bra{\phi^+_{t}}+\tfrac{1-\lambda_{time}}{2}(\ket{t_1}\bra{ t_1}^{\otimes 2}+\ket{t_2}\bra{ t_2}^{\otimes 2})),
\end{equation}
\end{widetext}
 where $\ket{\phi^+}\equiv(\ket{00}+\ket{11})/\sqrt{2}$ with $\ket{0}$ and $\ket{1}$ being computational basis states in the given DOF. We implement decoherence in time-bin qubits by slightly shifting in time the timing modes relative to each other. In combination with the measurement method, this functions as an approximation of decoherence~\footnote{Note, because the state could be made pure again by delaying one pulse relative to the other to achieve total overlap, the effective decoherence is actually reversible. However, this is often true of dephasing-based decoherence: in \textit{principle}, it could be undone if one knew the correct `anti-phase' fluctuations to apply.}. Because these measurements already involved adding noise to a maximally entangled state, effort was not made to perfectly tune the polarization state, so $\lambda_{pol}\approx0.9$ for these measurements.  As seen in Fig.~\ref{visdep}, we were able to observe an $I^{18}_{4422}$ Bell parameter over the qubit-entangled limit while the temporal visibility was above 0.75; in addition, a Bell violation excluding all local realistic models was observed for temporal visibilities of above 0.53, both in agreement with theory after accounting for our produced input state. 
 
\section{Remote Steering of Hyperentangled States}
Next, we verified the ability of one photon in the pair to ``steer'' the other photon using a two-basis steering scheme.  At a high level, quantum steering can be understood as a game in which Alice's goal is to convince Bob that she has distributed to him half of an entangled state. She does this by allowing him to measure his particle in a basis of his choice which he reports back to her; based on this information, she then measures her particle in a corresponding basis and reports the result back to Bob. Only if the correlations violate a steering inequality will Bob be convinced of the entanglement in the original state~\cite{RevModPhys.92.015001}.  
To make this more precise, suppose Alice and Bob share a bipartite state $\rho^{AB}$.  In the simplest steering test, Bob chooses between a pair of two-outcome projective measurements labeled by index $y\in\{1,2\}$.  The collection of Alice's post-measurement states is described by the assemblage $\{\rho_{b|y}\;:\;b=\pm 1, y=1,2\}$, for which we can write the steering inequality \cite{Cavalcanti-2016a}
\begin{equation}
\label{Eq:steering-ineq-dsummain}
 	S_{\text{Str}}\equiv\frac{1}{2}\left(\tr\left[F_0 X_1\oplus X_2\right]+\tr\left[F_1Z_1\oplus Z_2\right]\right) \leq\frac{1}{\sqrt{2}},
\end{equation}
where $F_y=\rho_{+1|y}-\rho_{-1|y}$ are formed from the untrusted assemblage and $\{X_1,Z_1\}$ and $\{X_2,Z_2\}$ are Pauli observables measured on orthogonal qubit subspaces, given $X$ and $Z$ are Alice's Pauli observables in the $\hat{x}$ and $\hat{z}$ directions of a qubit space (see Appendix~\ref{sec:steerder} for steering inequality derivation). The maximal value of $S_{\steer}$ is $1$.

To apply Eq.~\eqref{Eq:steering-ineq-dsummain} to our hyperentangled setup, we first generate the state $\ket{\Phi_4}=\frac{1}{2}(\ket{00}+e^{i\phi_r}\ket{11}+e^{i\phi_r}\ket{22}-\ket{33})$, with the computational basis states representing the same polarization-time states as Eq.~\eqref{psi4}, where $\phi_r$ is an uncalibrated random phase which the steering measurements are insensitive to. Note, the sign difference on the last term ensures that the state is not factorizable with respect to the two subsystems (one for each degree of freedom).  Let $X_1$ and $Z_1$ denote Pauli observables in the qubit subspace $\mc{H}_1=\text{span}\{\ket{Ht_1},\ket{Vt_2}\}$, and likewise $X_2$ and $Z_2$ denote Pauli observables in the qubit subspace $\mc{H}_2=\text{span}\{\ket{Ht_2},\ket{Vt_1}\}$.  In the steering protocol, Bob measures either $X_1\oplus X_2$ or $Z_1\oplus Z_2$, in each case obtaining a $\pm 1$ outcome.  Because of the symmetric nature of their shared state, Alice's optimal strategy is to measure the same observable as Bob. 
We achieved a steering parameter of 0.93$\pm$0.02$>1/\sqrt{2}$, thereby indicating Alice's ability to steer Bob's measurement and the entangled nature of the shared state.  Accounting for previous measurements of the entangled state's quality~\cite{chapman2020time}, we would predict this system to produce a steering parameter of 0.95, in good agreement with the value measured.

We then investigated the robustness of the steering parameter to Werner-state-like entanglement~\cite{PhysRevA.40.4277}. Specifically, we produced states of the form 			
\begin{widetext}
\begin{equation}
\rho_4 = \lambda\ket{\Phi_4}\bra{\Phi_4}+\frac{1-\lambda}{4}\bigg((\op{H}{H}^{\otimes 2}+\op{V}{V}^{\otimes 2}) \otimes (\op{t_1}{t_1}^{\otimes 2}+\op{t_2}{t_2}^{\otimes 2})\bigg),
\label{rho4}
\end{equation}
\end{widetext}
whose steerability bound for $\rho_4$ is $\lambda>\frac{1}{2}$ \cite{Augusiak-2015a}. We remark that the noisy state in Eq. \eqref{rho4} differs from standard Werner states that mix a maximally entangled state with white noise.  In contrast, $\rho_4$ reflects dephasing noise applied to $\ket{\Phi_4}$ that still maintains classical correlations between the two photons.  This state also differs slightly from the one generated in the previous Bell measurements above in that the degree of purity in the timing and polarization modes are here intended to be the same. In order to achieve this, simultaneous noise had to be added to both the time-bin and polarization degrees of freedom.  In the polarization degree of freedom, the $\lambda_{pol}$ parameter was tuned by inserting quartz crystals of varied thicknesses into the 810-nm path after the source, leading the two polarization components to walk off temporally from each other; alternatively, this can be interpreted as a frequency-dependent birefringent phase shift --- tracing over the 0.4-nm full-width half maximum spectral bandwidth of the 810-nm photons leaves the pairs in a partially mixed state. $\lambda_{time}$ was set in the same manner as for the four-setting Bell tests, by unbalancing the interferometers relative to each other~\footnote{For convenience, we used the measurement interferometers to impose these shifts; however, the result is completely equivalent to having such decoherence induced by the communication channel used to transmit the photons.}.  Special care must be taken when applying this method in conjunction with polarization decoherence techniques using birefringence, because both methods rely on displacing modes relative to each other in time; if polarization and timing modes are coupled (as they are in our particular measurement system), it is possible for these effects to act in opposite directions, leading to reduced effective decoherence in both modes. 

\begin{figure}
\centerline{\includegraphics[width=1\columnwidth]{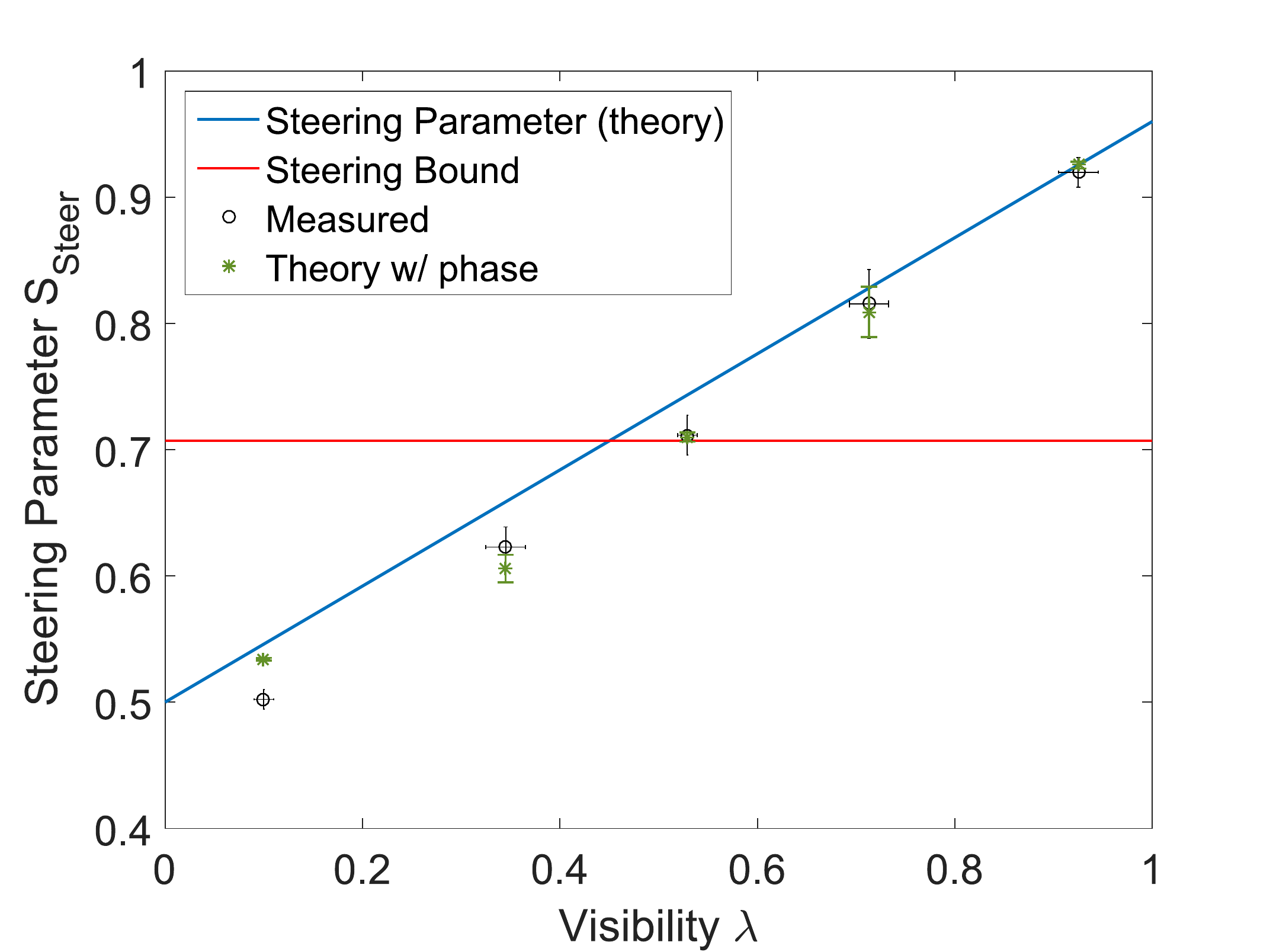}}
	\caption[Measured steering parameter values]{Measured steering parameter values.  The blue theory curve assumes a state exactly of the form given in Eq.~\eqref{rho4}, while the ``theory with phase'' predictions accounts for the entanglement phase Eq.~\eqref{phase}, as determined from a subset of the measurements. The error bars represent one standard deviation, assuming Poissonian counting statistics for the theory and was calculated using 5 measured samples for the measured data. }
	\label{steerplot}
	
\end{figure}

Because the quartz elements produce discrete changes in $\lambda_{pol}$, we tuned the path length difference of Alice's measurement interferometer to produce a value of $\lambda_{time}$ in the time mode that closely matched that of the polarization. Note that the phase of the complete entangled state depends on the relative path length differences of the interferometers: $\ket{\Phi_4^{\phi}}=\frac{1}{2}(\ket{00}+e^{i(\phi_{e_1}+\phi_r)}\ket{11}+e^{i\phi_r}\ket{22}+e^{i\phi_{e_2}}\ket{33})$. Hence the phases of the entangled state had to be readjusted to match $\ket{\Phi_4}$ for each noise value measured, using liquid crystals after the PBS of the measurement interferometer. This readjustment process becomes more challenging for small values of $\lambda_{time}$ (mostly mixed states), as the phase-insensitive noise dominates, leading to slightly different values for the entanglement phases at each noise level, despite efforts to re-tune this phase to zero between measurements.  In order to accurately predict the steering parameter observed at each level, it was therefore necessary to take into account this phase variation. Thus, for each noise level, we used the steering measurements to determine the entanglement phase    
\begin{equation}
  \phi_{e_2}=\arccos(4(p_2-p_1)),
  \label{phase}
  \end{equation}
where $p_1$ is the coincidence probability of Alice measuring -1 and Bob measuring +1 in basis 2, and $p_2$ is the probability of Alice and Bob both measuring -1 in basis 2.  Note that the measurements are insensitive to $\phi_{e_1}$ and $\phi_{r}$.

 As seen in Fig.~\ref{steerplot}, our steering measurement results show that the effects of this phase re-tuning were significant, as a model assuming this phase was zero does not accurately predict the observed steering parameter.  When this phase was included, the model was able to accurately predict the observed results. These results are the first to demonstrate quantum steering over multiple degrees of freedom using hyperentanglement.   Quantum steering generally provides a more robust method for certifying entanglement than violating some Bell inequality.  Our results demonstrate this fact for hyperentangled photons.  In particular, for $\lambda<0.65$ the state $\rho_4$ cannot violate any Bell inequality using projective measurements \cite{Acin-2006a, Augusiak-2015a}, whereas we were still able to verify its entanglement for part of this region via quantum steering
  until $\lambda<0.5$.

\section{Conclusion}
In this paper, we have shown a number of ways to detect and quantify entanglement in a hyperentangled photonic state.  The use of Bell tests beyond the CHSH inequality gives a device-independent indication of the higher-dimensional nature of the generated state with fewer measurements compared to a full state tomography. It also provides more reliable certification of higher-dimensional entanglement than other methods of entanglement certification/verification~\cite{Friis2019,PhysRevA.101.032302}, at the cost of more measurements.
This new feature could be more useful with increased study on the theoretic bounds of these inequalities. For example, our present work only shows that observing a sufficiently large Bell parameter provides evidence that the state is more than qubit-entangled, instead of indicating a specific dimension; if the spectrum of maximum Bell parameters as a function of entangled state dimension were known, then a more precise conclusion about the state dimension could be reached.  It is an open, interesting, and relevant theoretical question whether, and to what extent, diagnostic metrics like Bell and steering violations mirror the usability of the imperfect states for quantum information applications.  For example, we found that our intentionally decohered ququart state could not achieve a beyond-qubit violation when the visibility in the time degree of freedom was $<0.75$.  Are these also the values, e.g., at which the benefit from using higher-dimensional states for quantum error correction vanishes?  If so, then these metrics may play a key role in monitoring the performance capabilities of elements in a quantum network.

Further investigation into these less commonly measured Bell inequalities could thus lead to improved state characterization techniques.  Further, we demonstrated the ability to steer hyperentangled states, which could potentially enable a broad range of quantum applications that involve a trusted measurement system, such as MDI-QKD. To the extent that higher dimensional quantum states are found to be (in)valuable resources for advanced quantum protocols, nonlocality characterization methods such as those presented here enable more efficient state characterization and dimensionality estimation, than a full quantum state tomography, which has an exponentially increasing number of measurements for increasing state dimension.

\acknowledgements{Thanks to Michael Wayne and Kristina Meier for discussions regarding design of the time-bin sorting circuit, and to Chris Chopp for assistance in its prototyping and printed-circuit-board layout. Thanks to Kelsey Ortiz for graphics assistance in creating the final version of the experimental setup figures. C.K.Z., J.C.C., and P.G.K. acknowledge support from NASA Grant No. NNX13AP35A and NASA Grant No. NNX16AM26G. J.C.C. acknowledges support from a DoD, Office of Naval Research, National Defense Science and Engineering Graduate Fellowship (NDSEG) and from the U.S. Department of Energy, Office of Science, Office of Advanced Scientific Computing Research, through the Transparent Optical Quantum Networks for Distributed Science Program (Field Work Proposals ERKJ355). This work was partially performed at Oak Ridge National Laboratory (ORNL). ORNL is managed by UT-Battelle, LLC, under Contract No. DE-AC05-00OR22725 for the DOE. P.G.K and E.C. are partially supported by the National Science Foundation (NSF) Quantum Leap Challenge Institute for Hybrid Quantum Architectures and Networks, Award No. 2016136.}

\appendix

\section{Verification with CHSH inequality} \label{sec:CHSH}
For reference, we first measured a CHSH inequality~\cite{PhysRevLett.23.880} on each of the degrees of freedom (DOF) individually.  The CHSH Bell parameter can be calculated from the quantum correlations between measurement results made in different bases for the two photons:
			\begin{equation}
 	S=E(a,b)+E(a',b)+E(a',b)-E(a',b').
 \end{equation}
 For local realistic states, $|S|\leq 2$ and for quantum states, $|S|\leq 2\sqrt{2}$~\cite{cirel1980quantum}. E($a$,$b$) can be calculated from the coincidences between Alice and Bob's detector pairs when Alice measures in basis $a$ and Bob measures in basis $b$:
			\begin{equation}
 	E(a,b)\equiv\frac{N_{11}+N_{22}-N_{12}-N_{21}}{N_{11}+N_{22}+N_{12}+N_{21}}.
 \end{equation}
 Here $N_{ij}$ is the number of events for which Alice's measurement outcome was $i$ and Bob's measurement outcome was $j$.  For the polarization DOF measurements, outcome $1$ ($2$) corresponded to detections on Detectors 1 or 2 (3 or 4) in Fig~\ref{setupfig}; for the timing DOF measurements, $1$ ($2$) corresponded to detections on Detectors 1 (4).
 
For these measurements, the entanglement source was adjusted to display entanglement in only one DOF at a time. To generate photon pairs that were only entangled in polarization, the long arm of the pump interferometer was blocked so that the nonlinear crystal was driven by only one time mode. We also adjusted the pump half-wave plates to optimize the state for the CHSH measurements, i.e., minimizing  $\bra{DA}$ coincidences~\footnote{The first half wave plate was rotated normally (about the propagation axis) to change the probability between $\ket{H}$ and $\ket{V}$ and the other wave plate was tilted about the vertical axis, with the slow/fast axes in the H/V basis, to change the phase between $\ket{H}$ and $\ket{V}$.}.  To measure pairs entangled only in time bin, we rotated the first pump half-wave plate so that mostly the clockwise process in the Sagnac source was activated and most of the downconversion photons had a definite horizontal polarization. We also inserted a half-wave plate and polarizer on Bob's side to project the polarization onto $\bra{H}$ and then to rotate it to $\bra{D}$ since the analyzer projects onto states like $(\bra{H t_1}\pm\bra{V t_2})/\sqrt{2}$ and $(\bra{V t_1}\pm\bra{H t_2})/\sqrt{2}$.  We then carried out a Bell measurement using the optimal measurement settings for a maximally entangled state ($a=0^\circ$, $a'=45^\circ$, $b=22.5^\circ$, $b'=67.5^\circ$)~\cite{PhysRevLett.49.1804}.   For the polarization basis-measurements, these values were set using the half-wave plates before each measurement interferometer; for the time-bin basis measurements, they were set by the half-wave plates after each measurement interferometer. We observed a Bell parameter of 2.58$\pm$0.02 for the polarization-entangled state and 2.40$\pm$0.02 for the time-bin entangled state.  From earlier measurements on the quality of the source entanglement~\cite{chapman2020time}, we would expect a polarization Bell parameter of 2.75 and a time-bin Bell parameter of 2.68.  The lower observed values are explainable by imperfect phase tuning and imperfect PBS extinction.  The time-bin value is further lowered due to residual phase averaging caused by path length fluctuations in the interferometers, as well as slight static mismatches between the three unbalanced interferometers.

\section{Bell Inequality Optimization Analysis} \label{sec:BIAO}

Like the standard Bell parameter, the $I_{4422}$ inequalities are a synthesis of three types of probabilities. The first type of probability is the chance that Alice observes a particular measurement outcome (e.g., 1) when she measures in a particular basis (e.g., $a$). Similarly, the second type of probability is the chance that Bob observes a particular measurement outcome when he measures in a particular basis.  The final type of probability is the chance that Alice and Bob both measure particular outcomes when they measure in particular bases.  The inequality then consists of a set of coefficients for these probabilities, such that the sum of the probabilities multiplied by their respective coefficients cannot exceed a certain limit in a local theory (e.g., $I^{18}_{4422}\leq 0$). In our measurement protocol, for simplicity we take outcome 1 to be a detection event on Detector 1, and outcome 2 to be a detection event on \textit{any} of the other three detectors. Of course, we could have assigned any detector to outcome 1, and the other three to outcome 2. 

\begin{table}
\caption{Numerically optimized settings for the $I^{18}_{4422}$ Bell inequality for the experimental setup in Fig.~\ref{setupfig}.  While the optimization was carried out over all HWP and QWP settings, for the optimum, all QWP settings were 0$^\circ$. With these settings, the system can achieve a Bell parameter of up to 0.46, while a local model is limited to a value of 0, and a qubit-entangled state is limited to 0.18~\cite{4Dineq,P_l_2008}.}
\label{bellsettab}
\begin{center}
	\begin{tabular}{ c c c c c}
	\hline
	\hline
		  & Basis 1 & Basis 2 & Basis 3 & Basis 4 \\ \hline

		Alice HWP 1 & $45 ^{\circ}$  & $24^{\circ}$  & $58^{\circ}$  & $8^{\circ}$ \\ 
		Alice HWP 2 & $12^{\circ}$  & $43^{\circ}$  & $22^{\circ}$  & $15^{\circ}$  \\ 
		Bob HWP 1 & $12^{\circ}$  & $42^{\circ}$  & $7^{\circ}$  & $-33^{\circ}$ \\ 
		Bob HWP 2 & $20^{\circ}$  & $49^{\circ}$  & $22^{\circ}$  & $15^{\circ}$ \\ 
	\hline
	\hline		
			\end{tabular}
\end{center}
\end{table}

 Our measurement system cannot project onto an arbitrary ququart state (e.g., we cannot project onto $\ket{H}\otimes(\ket{t_1}+\ket{t_2})/\sqrt{2}$ because the two time-modes always have orthogonal polarizations in our system), thus it is not possible to obtain a violation for some of the inequalities, and for those with a violation, the maximum violation achievable with our system does not necessarily reach the theoretic limit of the inequality.  For this reason, we explored all known $I_{4422}$ inequalities, and not just those with the largest theoretic difference for qubit and qutrit states.  
 
After choosing the inequality,
\begin{widetext}
\begin{align}
    I^{18}_{4422}&=2\text{Pr}(a_1,b_1)+2\text{Pr}(a_1,b_2)+2\text{Pr}(a_1,b_3)-\text{Pr}(a_1,b_4)\notag\\
    &+2\text{Pr}(a_2,b_1)+\text{Pr}(a_2,b_2)-2\text{Pr}(a_2,b_3)+2\text{Pr}(a_2,b_4)\notag\\
    &+2\text{Pr}(a_3,b_1)-2\text{Pr}(a_3,b_2)-2\text{Pr}(a_3,b_3)-2\text{Pr}(a_3,b_4)\notag\\
    &-\text{Pr}(a_4,b_1)+2\text{Pr}(a_4,b_2)-2\text{Pr}(a_4,b_3)-\text{Pr}(a_4,b_4)\notag\\
    &-2\text{Pr}(a_1)-2\text{Pr}(a_2)-2\text{Pr}(b_1)-2\text{Pr}(b_2)\leq 0,
\end{align}
\end{widetext} 
we optimized the experimental measurement settings for the best violation by our system with that inequality; the optimal settings are in Table~\ref{bellsettab}.
 
\section{Steering Inequality Proof} \label{sec:steerder}
Suppose Alice and Bob share a bipartite state $\rho^{AB}$.  In the simplest steering test, Bob chooses between a pair of two-outcome projective measurements, with measurement $y\in\{1,2\}$ described by orthogonal projectors $\{B_{+1|y},B_{-1|y}\}$.  The collection of Alice's post-measurement states is described by the assemblage $\{\rho_{b|y}\;:\;b=\pm 1, y=1,2\}$, where
\begin{equation}
\label{Eq:assemblage}
\rho^A_{b|y}=\tr_B\left[\mbb{I}\otimes B_{b|y}\rho^{AB}\right]
\end{equation}
(in general, any collection of positive operators $\{\rho_{b|y}\}_{b,y}$ such that $\sum_{b=\pm 1}\tr[\rho_{b|y}]=1$ for all $y$ is called a state assemblage).

If $\rho^{AB}$ is not entangled, then it can be expressed in separable form as $\rho^{AB}=\sum_{\lambda}p(\lambda)\op{\alpha_\lambda}{\alpha_\lambda}^A\otimes\op{\beta_\lambda}{\beta_\lambda}^B$.  Substituting this into Eq.~\eqref{Eq:assemblage} yields
\begin{equation}
\label{Eq:assemblage2}
\rho_{b|y}=\sum_\lambda p(b|y,\lambda)\op{\alpha_\lambda}{\alpha_\lambda}^A,
\end{equation}
where $p(b|y,\lambda):=\bra{\beta_\lambda}B_{b|y}\ket{\beta_\lambda}p(\lambda)$.  Any state assemblage that can be expressed in the form of Eq.~\eqref{Eq:assemblage2} is known to satisfy a \textit{local hidden state} (LHS) model.  Note that measuring one-half of a separable state always yields an LHS assemblage, but the converse is not true~\cite{PhysRevLett.98.140402}.

This discussion shows that every non-entangled state generates an LHS assemblage when measured on one side.  Conversely, if the post-measurement assemblage does not satisfy a LHS model, then the original state $\rho^{AB}$ was entangled.  This method of entanglement detection is ``semi-device-independent'' since the only trusted device is Alice's measurement apparatus in testing the generated assemblage $\{\rho_{b|y}\}_{b,y}$.  For the simplest case of two dichotomic measurement choices on Bob's side, it has been shown in Ref.~\cite{Cavalcanti-2016a} that all assemblages satisfying a LHS model obey the steering inequality

\begin{equation}
\label{Eq:steeringineqSM}
\frac{1}{2}\left(\tr\left[F_0 X\right]+\tr\left[F_1Z\right]\right) \leq\frac{1}{\sqrt{2}},
\end{equation}
where $F_y=\rho_{+1|y}-\rho_{-1|y}$ are formed from the untrusted assemblage, and $X$ and $Z$ are Alice's Pauli observables in the $\hat{x}$ and $\hat{z}$ directions of her qubit space.  If $\{X_1,Z_1\}$ and $\{X_2,Z_2\}$ are Pauli observables measuring on orthogonal qubit subspaces, then the previous inequality can be extended to read
\begin{equation}
 	S_{\steer}:=\frac{1}{2}\left(\tr\left[F_0 X_1\oplus X_2\right]+\tr\left[F_1Z_1\oplus Z_2\right]\right) \leq\frac{1}{\sqrt{2}}.
\label{Eq:steeringineqdsumSM}
\end{equation}
We prove this inequality by first noting that
\begin{align}
\tr[F_0X_1\oplus X_2]&=\tr[\Pi_1 F_0 \Pi_1 X_1]+\tr[\Pi_2 F_0 \Pi_2 X_2]\\
\tr[F_1Z_1\oplus Z_2]&=\tr[\Pi_1 F_1 \Pi_1 X_1]+\tr[\Pi_2 F_1 \Pi_2 X_2],
\end{align}
where $\Pi_i$ is a projector onto the $\{X_i,Z_i\}$ qubit subspace.  Since $F_y=\tr_B\left[\mbb{I}\otimes (B_{+|y}-B_{-|y})\rho^{AB}\right]$, we have
\begin{align}
 \Pi_iF_y\Pi_i&=p_i\tr_B\left[\mbb{I}\otimes (B_{+|y}-B_{-|y})\rho_i^{AB}\right]\notag\\
 &=:p_iF_{i,y},
\end{align}
where $\rho_i^{AB}=(\Pi_i\otimes\mbb{I})\rho^{AB}(\Pi_i\otimes\mbb{I})/p_i$ is a normalized state with $p_i=\tr[(\Pi_i\otimes\mbb{I})\rho^{AB}]$.  Therefore,
\begin{align}
\tr[&F_0X_1\oplus X_2]+\tr[F_1Z_1\oplus Z_2]\notag\\
&=p_1(\tr[X_1 F_{1,0}]+\tr[Z_1 F_{1,1}])\notag\\
&\quad+p_2(\tr[X_1 F_{2,0}]+\tr[Z_1 F_{2,1}])\notag\\
&\leq \sqrt{2},
\end{align}
by Eq.~\eqref{Eq:steeringineqSM} above.

%

\end{document}